\documentclass[12pt]{article}
\usepackage{amssymb,amsmath}
\def\d{{\partial}}
\def\um{\frac{1}{2}}
\def\A{{\cal A}}
\def\X{{\cal X}}
\def\N{{\cal N}}
\def\Tr{{\rm Tr}}
\def\P1{{\bf P}^1}
\def\ln{{\rm ln}}
\def\0{{\nonumber}}
\def\a{\begin{eqnarray}}
\def\b{\end{eqnarray}}
\def\be{\begin{equation}}
\def\ee{\end{equation}}
\begin{document}

\begin{titlepage}

\hfill SISSA/11/2006/EP

\hfill hep-th/0603083

\vspace{1cm}
 
\centerline{\huge{Flavour from partially resolved singularities}}
\vspace{1.5cm}

\centerline{\large{G.~Bonelli, L.~Bonora and A.~Ricco}}
\vspace{.5cm}

\centerline{International School of Advanced Studies (SISSA) and INFN, Sezione
  di Trieste}
\centerline{ via Beirut 2-4, 34014 Trieste, Italy}

\begin{abstract}
In this letter we study topological open string field theory on D--branes
in a IIB background given by non compact CY geometries 
${\cal O}(n)\oplus{\cal O}(-2-n)$ on $\P1$ with a singular point
at which an extra fiber sits. We wrap
$N$ D5-branes on $\P1$ and $M$ effective D3-branes at singular points,
which are actually D5--branes wrapped on a shrinking cycle. 
We calculate the holomorphic Chern-Simons partition function for the above models 
in a deformed complex structure and find that it 
reduces to multi--matrix models with flavour.
These are the matrix models whose resolvents have been shown to satisfy the 
generalized Konishi anomaly equations with flavour.
In the $n=0$ case, corresponding to a partial resolution of the $A_2$ singularity, 
the quantum superpotential in the ${\cal N}=1$ unitary SYM 
with one adjoint and $M$ fundamentals is obtained.
The $n=1$ case is also studied and shown to give rise to two--matrix 
models which for a particular set of couplings can be exactly solved. 
We explicitly show how to solve such a class of models by a quantum
equation of motion technique.
\end{abstract}

\end{titlepage}

\section{Introduction}

A type II background of the form $\mathbb{R}^{1,3} \times X$, 
where $X$ is Calabi-Yau threefold, containing BPS branes 
and fluxes, generically produces low energy effective theories with $N=1$ 
supersymmetry. 
While the relation between open/closed string moduli and effective gauge 
theories (``geometric engineering'')
is quite well understood in the particular case of $N=2$ supersymmetry, 
the $N=1$ case still lacks a complete understanding.
For this reason, the study of the dynamics of branes in Calabi-Yau manifolds 
has attracted a lot of attention both for its theoretical and phenomenological 
applications, e.g. \cite{geometric}. 
For instance, considering D--branes 
wrapped around two--cycles in a non--compact CY one can link the superpotential of 
the $N=1$ supersymmetric gauge theories living on the space--filling
branes to the deformation of the CY geometry \cite{Kachru:2000ih, AK}. 

In previous papers \cite{BBR,Proc}, building up on the existing literature 
(see e.g. \cite{Ferrari:2003vp} and references therein), we analyzed 
type IIB superstring theory background with space filling D5--branes 
wrapped around 
two--cycles of a non--compact Calabi--Yau threefold.
We 
explicitly
showed that, upon topological twist, the theory reduces to a (multi--)matrix
model whose potential describes deformations of complex structures in the
singular Calabi--Yau threefolds. Some geometrical properties of these spaces have been 
studied in \cite{2}.
In this paper we 
extend this analysis by including flavour. Phrasing it in another way, we 
geometric engineer $N=1$ 
(supersymmetric gauge) theories containing fields in the fundamental 
representation of the gauge group, by introducing a suitable
brane background.
 
More precisely we study topological open string field theory 
on branes in a IIB background (for the relation between topological
strings and superstrings in this context, see \cite{Vafa:2000wi}),
where $X$ is a non--compact CY given by 
${\cal O}(n)\oplus{\cal O}(-2-n)$ on $\P1$ with a singular point
at which an extra fiber sits.
We wrap $N$ space--filling D5-branes on 
$\P1$ and complete the configuration with $M$ `effective' D3-branes stuck at 
the singular point. In the case $n=0$ we show that
the latter can be actually interpreted as D5--branes wrapped
around a two--cycle which is subsequently shrunk to the singular point.
This requires a partially resolved geometrical set up, which we describe
in detail (see the Appendix). While the
D5--brane sector gives rise, as in the smooth case, to the
superpotential (and, from a geometrical point of view,
describes deformations of the smooth CY complex structure),
the effective D3-brane sector 
gives rise to a novel part of the spectrum, the corresponding superpotential data being
related to {\it linear} deformations of the CY complex structure along the extra fiber.

We calculate the partition function for the above models 
by direct computation of the holomorphic Chern-Simons partition function
in a deformed complex structure and find that it 
reduces to multi--matrix models with flavour \cite{Argurio:2002xv, McGreevy:2002yg, after}.
These are the matrix models whose resolvents have been shown to satisfy the 
generalized Konishi anomaly equations with flavour \cite{seiberg}.
In the $n=0$ case, the quantum superpotential in the ${\cal N}=1$ $U(N_c)$ 
gauge theory with one adjoint and $N_f$ fundamentals is obtained.
The $n=1$ case is studied in detail and shown to give rise to matrix 
models which for a particular set of couplings turn out to be solvable. 
In general the flavour can be integrated out and one obtains a matrix model 
with a polynomial plus a logarithmic potential. We give explicit examples 
of how to solve the latter with the technique of the quantum equations of
motion.
For recent related works, see \cite{1}.
 
The paper is organized as follows. 
In section 2 we study the reduction of the holomorphic Chern-Simons 
on the vector bundle ${\cal N}={\cal O}(n)\oplus{\cal O}(-2-n)$ on $\P1$, 
augmented by an extra fiber, to 
a two dimensional theory over $\P1$.
In section 3 we show that the calculation of the relevant partition function
reduces to (multi)matrix integrals with flavour, and produce some explicit
examples. 
In section 4 we study and solve a specific class of examples of 
matrix models arising in the above set-up.
Section 5 is left for few concluding remarks. In Appendix we work out 
in detail the geometry of of a partially resolved $A_2$ singularity which is
relevant for the $n=0$ case.

\section{Holomorphic Chern-Simons and two--dimensional\\ gauge theories}

Let us consider
the non compact CY geometries given by the total space of the vector bundle
${\cal N}={\cal O}(n)\oplus{\cal O}(-2-n)$ on $\P1$ augmented by an extra fiber
at a singular point of the $\P1$ and let us denote this space
\footnote{More general 
geometries built as the total space of a rank 2 holomorphic vector bundle over 
a generic Riemann surface could be considered 
along the lines of \cite{BBR}, but we will not do it here.}
by $CY_n$.
For $n=0$ this space is the partial resolution of an $A_2$ singularity, as it 
is described in detail in the Appendix. The other cases with $n>0$ are a 
generalization thereof.

We consider type IIB theory on $R^{1,3}\times CY_n$ with 
$N$ D5-branes along $R^{1,3}\times \P1$ 
and $M$ 3-branes along $R^{1,3}$, the latter being stuck at a singular point 
in $\P1$ where the extra fiber sits.
The Calabi--Yau threefold can be topologically written as 
${\cal N} \vee \mathbf{C}^2$, 
where $\vee$ is the reduced union, i.e. the disjoint union of two spaces 
with a base point of each 
identified.
 In the case $n=0$ we interpret this 
geometric background in the following way. We start with a resolution of an
$A_2$ singularity, see Appendix, and 
wrap $N$ D5--branes on one
cycle and $M$ and the other. When we blow down the latter, the  
$M$ D5--branes wrapped around the shrunk cycle will appear
as effective D3--branes stuck at the singular point in the remaining 
${\bf P}^1$: they cannot vibrate along the base, 
while they are free to oscillate along the extra fiber\footnote{In this paper
we are using a simple (and probably too poor) language 
in which B-branes are complex submanifolds of the target space. 
In a more sophisticated approach isolated fibers over singular points should 
be replaced by skyscraper sheaves and, in general, B--branes should be embedded
as complexes in derived categories of coherent sheaves, see \cite{Aspinwall} for a review.
We show that our approach nevertheless captures some relevant topological 
information.}. As we said, for generic $n$ we single out a point on $\mathbf{P}^1$ and 
add an extra fiber to render it singular, but the interpretation as blow-down of
a smooth cycle is not as evident as for $n=0$.

In this paper we would like to show that,
upon topological twist, the superpotential of this theory can be calculated 
by means of the second quantized topological string theory.

The latter is given by the holomorphic Chern-Simons theory \cite{Witten:1992fb} 
(see also the lectures \cite{Marino:2004eq})
\begin{equation}
S(\A_T)=\frac{1}{g_s^2}\int_{CY_n} 
\Omega\wedge Tr_{N+M}\left(\frac{1}{2} \A_T\wedge
\bar\partial \A_T + \frac{1}{3} \A_T \wedge \A_T \wedge \A_T\right)
\label{hCS}
\end{equation}
where $\A_T\in T^{(0,1)}\left(CY_n\right)$ with 
full Chan-Paton index $N+M$. The total string field $\A_T$ can be expanded as
$$
\A_T=\left(\begin{matrix}\A & \X \\ \tilde \X & 0\end{matrix}\right)
$$
where $\A$ is the string field for the 5-5 sector and $(\X,\tilde \X)$ for 
the 5-3 and 3-5
open strings. The 3-3 sector is irrelevant to us (anyway, see next footnote).

The action (\ref{hCS}) reduces to
$$
S(\A_T)=\frac{1}{g_s^2}\int_{CY_n} 
\Omega\wedge \left[Tr\left(\frac{1}{2} \A\wedge
\bar\partial \A + \frac{1}{3} \A \wedge \A \wedge \A\right)\right.
$$
\begin{equation}
\left.
+\frac{1}{2}\sum_{I=1}^M\left(\bar D_\A\tilde\X_I \wedge \X^I +\tilde\X_I \wedge \bar D_\A \X^I\right)\right] 
\label{rhCS}
\end{equation}
where gauge indices are not explicitly shown, $I=1,\dots,M$ is the flavour index and 
$D_\A$ is the covariant derivative in the (anti-)fundamental representation.

The reduction of the open string field to the D5-brane world-volume ${\bf P}^1$ 
is obtained via an auxiliary  invertible bilinear form on 
${\cal N}\otimes\bar{\cal N}$ which we denote by $K$ and its associated Chern connection
$\Gamma_{\bar z}=K^{-1}\partial_{\bar z}K$.
The reduction condition for the 5-5 sector is
$(D_\Gamma\A)^{\cal N}=0$, where $D_\Gamma$ is the covariant derivative w.r.t. $\Gamma$
and the $\N$ index denotes projection along the fiber directions.
The reduction conditions for the 5-3 and 3-5 sector read 
$(D_\Gamma\X)^{\cal N}=0$ and $i_{\partial_{\bar z}}\X|_{{\bf P}^1}=0$,
as well as 
$(D_\Gamma\tilde\X)^{\cal N}=0$ and $i_{\partial_{\bar z}}\tilde\X|_{{\bf P}^1}=0$
.
The last condition specifies that the D3-branes are stuck at the singular point in
$\P1$
and therefore their oscillations along $T\P1$ are inhibited
\footnote{Actually, the 3-3 sector would appear as the $C$ component in
$
\A_T=\left(\begin{matrix}\A & \X \\ \tilde \X & C\end{matrix}\right)
$.
It would modify the action by $\Delta S=\frac{1}{g_s^2}\int_{CY_n} 
\Omega\wedge \left[Tr_{M}\left(\frac{1}{2} C\wedge
\bar\partial C + \frac{1}{3} C \wedge C \wedge C\right)
+ \tilde X \wedge C\wedge X\right]$.
Upon the reduction condition for the 3-3 sector
$i_{\partial_{\bar z}}C|_{{\bf P}^1}=0$, we see that the last two terms
give vanishing contribution (neither $C$ nor the $X$'s have a 
$d\bar z$ component to 
complete a top-form on $CY_n$), therefore the 3-3 sector decouples.}.

The reduction of the 5-5 sector has been already discussed in \cite{BBR} and 
will not be repeated here. The reduction of the flavour sector can be 
carried out with the same technique.
Let $X_{\bar i}^I=K_{\bar i j}Q^{jI}$ and analogously
$\tilde X_{\bar iI}=K_{\bar i j}\tilde Q^j_I$, where
$\X=X_{\bar i}dw^{\bar i}$ and 
$\tilde \X=\tilde X_{\bar i}dw^{\bar i}$
solve the above reduction conditions.
The reduction of the Lagrangian density to the $\P1$ for the 5-3 sector
follows the same logic as for the 5-5 sector.
The proper pullback to the base is performed patch by patch 
with the help of $K$ as contraction of the 
hCS (3,3)-form Lagrangian by the two bi-vector fields 
$k=\frac{1}{2}
\epsilon_{ij}K^{i \bar l}K^{j \bar k}
\frac{\partial}{\partial \bar w^l}
\frac{\partial}{\partial \bar w^k}$ 
and $\rho=\frac{1}{2}
\epsilon^{ij}\frac{\partial}{\partial w^i}
\frac{\partial}{\partial w^j}$.

The resulting $(1,1)$-form Lagrangian density reads
\begin{equation}
L_{fl.red.}=
\um \epsilon_{ij}\tilde Q^i_I D_{\bar z} Q^{jI}
-
\um \epsilon_{ij}D_{\bar z}\tilde Q^i_I Q^{jI}
\label{Lrf}\end{equation}
which is independent on $K$. This proves that our reduction mechanism is well defined.
The action (\ref{Lrf}) was given in \cite{Witten:2003nn} in a similar context
and is always a $\beta\gamma$-system.
Combining with the 5-5 sector, the total reduced action then reads
\begin{equation}
L_{red.}= 
\um \epsilon_{ij}\Tr \phi^i D_{\bar z}\phi^j
+
\um \epsilon_{ij}\tilde Q^i_I D_{\bar z} Q^{jI}
-
\um \epsilon_{ij}D_{\bar z}\tilde Q^i_I Q^{jI}
\label{Lr}\end{equation}

It is straightfoward to generalize the gauge fixing procedure for the 
5-5 sector, see \cite{Proc}, to the total system
to show that the partition function of the theory above reduces to 
matrix integrals over the $\partial_{\bar z}$ zero--modes
of the fields.

\section{Deformed complex structures and matrix models}

The deformation of the complex structure of the singular spaces
we are considering can be split in two operations, namely the 
deformation of the smooth part and the deformation of the extra fiber at 
the singular point.




As far as the smooth part is concerned,
the deformed complex structures to which we confine are of the form
obtained by glueing the north and south patches of the fibers above 
the sphere as
\begin{equation}
\omega_N^1=z_S^{-n}\omega_S^1 ,
\quad{\rm and}\quad
\omega_N^2=z_S^{2+n}\left[\omega_S^2+\partial_{\omega^1}B\left(z_S,
\omega_S^1\right)\right]
\label{varfin}\end{equation}
This, as is well--known, preserves the Calabi-Yau property of the six manifold.
As widely discussed in \cite{BBR} the glueing conditions (\ref{varfin})
gets promoted to the glueing conditions for the 5-5 sector, that is the 
chiral adjoints.

Now let us deal with the analogous deformation for the 
5-3 and 3-5 sectors. 
Let $\hat P$ be the point on $\P1$ where the extra fiber sits and let 
$(x^1,x^2)$ be the coordinates along the latter.
Before the complex structure deformation, the extra fiber glueing is 
given by
$$
x_N^1={\hat z_S}^{-n}x_S^1 ,
\quad{\rm and}\quad
x_N^2={\hat z_S}^{2+n}x_S^2.
$$
The complex structure deformations we confine to for this sector 
are linear and are described by the glueing conditions
\begin{equation}
x_N^1={\hat z_S}^{-n}x_S^1 ,
\quad{\rm and}\quad
x_N^2={\hat z_S}^{2+n}\left[x_S^2+M\left({\hat z}_S,\omega_S^1\right)x_S^1\right],
\label{varvec}\end{equation}
where $M$ is locally analytic on ${\bf C}\times\left(U_N\cap U_S\right)$.
This can be cast in the form
\begin{eqnarray}
M(z, \omega) = \sum_{d=1}^{\infty} \sum_{k=0}^{n d+2}  m_d^{k} z^{-k-1} 
\omega^d .
\end{eqnarray}
Notice however that only a subset of the parameters $m_d^{k}$ parameterize 
actual deformations of the complex structure, since only a part of them cannot 
be reabsorbed by a local reparametrization.

The deformed glueing condition (\ref{varvec}) is coupled to the 3-5 sector
of the topological open string field since the 3-branes only vibrate 
transversely along the extra-fiber.

Note that we are obtaining different string backgrounds on our geometry, by 
considering the smooth variety (\ref{varfin}) 
and `attaching' to it additional fibres, eq. (\ref{varvec}).
The function $M$ then generates the variation of the complex structure of 
the CY along the singular fiber.

This deformed geometry (\ref{varvec}) can be implemented in the reduction of 
the open string field in a way much similar to the one followed for the pure 
5-5 sector in the smooth case. 
This is done by promoting (\ref{varfin}-\ref{varvec}) to the glueing 
conditions of the reduced string field components $\phi^i$, $X^i$ and 
$\tilde X^i$ with a patch by patch singular field redefinition which reabsorbs
the deformation terms $(B,M)$.
In these singular coordinates the fields glue linearly and we can apply the 
results of the previous section, obtaining in this way the Lagrangian 
(\ref{Lr}) in the singular field coordinates.
Going back to the regular coordinates, one gets the action
$$
S_{red.}=\int_{\P1} 
\Tr \left[\phi^2 D_{\bar z}\phi^1
+
\tilde Q^2_I D_{\bar z} Q^{1I}
+
Q^{I2} D_{\bar z} \tilde Q^1_I\right]
$$
\begin{equation}
+\oint_{aequator} dz \left[\Tr B(\phi^1,z)+ \tilde Q_I^1 M(\phi^1,z) Q^{1I}\right].
\label{Lrd}\end{equation}

The partition function of the latter theory can be calculated 
as in \cite{Proc} and one gets as a result a multi matrix--model of vector 
type, namely with additional interactions with flavours.
The fields $\phi^1$ and $(Q^1,\tilde Q^1)$ contribute only through
their $\partial_{\bar z}$
zero--modes. Under the above glueing prescription we expand
$\phi^1(z)=\sum_{i=0}^n z^i X_i$
and analogously
$Q^{1I}(z)=\sum_{i=0}^n z^i q^I_i$
and
$\tilde Q^1_I(z)=\sum_{i=0}^n z^i \tilde q_{iI}$, where $X_i$, $q^I_i$ and 
$\tilde q_{iI}$ are matrix
and vector constant coefficients.

Specifically the partition function, after the above calculations, reads
\begin{equation}
Z_n=\int \prod_{i=0}^n dX_i dq_i d\tilde q_i\,
{\rm e}^{\frac{1}{g_s^2}\left(\Tr {\cal W}(X) + \tilde q^I_i {\cal M}(X)_{ij} 
q_{Ij}\right)}
\label{vmmm}\end{equation}
where 
$${\cal W}(X)=\oint dz\, B\left(\sum_{i=0}^n z^i X_i,z\right)$$
is the 5-5 contribution already obtained in \cite{BBR}
and
$$
{\cal M}(X)_{ij}=\oint dz\, z^{i+j}M\left(\sum_{k=0}^n z^k X_k,z\right)
$$
represents the 5-3/3-5 coupling. 

We are therefore finding in the general case a multi matrix model with flavour symmetry.
As explained in \cite{review},
these matrix models 
are related to the quantum superpotential of a ${\cal N}=1$ SYM  
with
$(n+1)M$ chiral multiplets in the fundamental/anti-fundamental and $n+1$ in 
the adjoint representation.

Let us specify a couple of examples which turn out to be interesting.

\subsection{One chiral in the adjoint ($n=0$ case)}

In particular, if $n=0$ and the CY manifold is the total space of 
$\left[{\cal O}(-2)_{\P1}\vee {\bf C}\right]\times {\bf C} $ we have a single set of 
constant zero--modes.
Choosing 
$$
B(\omega^1,z)=\frac{1}{z}{\cal W}(\omega^1)
\quad {\rm and} \quad
M(\omega^1,z)=\frac{1}{z}{\cal M}(\omega^1)
$$
we find
\begin{equation}
Z_0=\int dX dq d\tilde q\,
{\rm e}^{\frac{1}{g_s^2}\left(\Tr {\cal W}(X) + 
\tilde q^I {\cal M}(X) q_{I}\right)}.\label{Z0}
\end{equation}

This partition function is then an extended matrix model with vectorial 
entries of the same type as the ones first considered in 
\cite{Argurio:2002xv}, which gives rise, in the
large $N$ expansion, to the quantum superpotential for ${\cal N}=1$ with 
$M$ flavour.
Actually a deeper analysis of these models started soon after \cite{after} 
culminated in \cite{seiberg}, where it was shown that the resolvent of the 
above enriched matrix model (\ref{Z0}) solves the generalized Konishi 
anomaly equations of the corresponding four dimensional gauge theory.
It is natural therefore to conjecture that the same is true for the 
other matrix models (\ref{vmmm}) that we have just obtained above.

In particular, if ${\cal M}(\omega^1)=\omega^1-m$ and 
${\cal W}'(\omega^1)=(\omega^1)^{N}+\dots$, 
the correct SW curve
$$
y^2=[{\cal W}'(x)]^2 +\Lambda^{2N-N_f}(x-m)^{N_f}
$$
is recovered \cite{McGreevy:2002yg}.
The above SW curve should be related to the deformed partially 
resolved geometry we are considering.

\subsection{Two adjoint chirals ($n=1$ case)}\label{31}

As a further example, let us discuss the result we obtain in the $n=1$ case.
Let us denote by $\Phi_0$ and $\Phi_1$ the two adjoint chiral superfields,
then the formulas for the superpotential and for the mass term read
\begin{eqnarray}
\mathcal{W}(\Phi_0, \Phi_1) &=& \sum_{d=1}^{\infty} \sum_{k=0}^{d} 
                           t^{k}_d 
                           \sum_{i_1, \dots i_d = 0,1 \atop i_1 + \dots + 
			   i_d =k} \Phi_{i_1} \dots \Phi_{i_d} \nonumber \\ 
\mathcal{M}(\Phi_0, \Phi_1) &=& \sum_{d=1}^{\infty} \sum_{k=0}^{d} 
          \left(\begin{array}{ll}
             m_d^{k} &  m_d^{k+1}  \\
             m_d^{k+1} & m_d^{k+2} 
          \end{array} \right)
                       \sum_{i_1, \dots i_d = 0,1 \atop i_1 + \dots + i_d =k} 
		       \Phi_{i_1} \dots \Phi_{i_d} \ . 
\end{eqnarray}

Note, for later use, that it is possible to produce a superpotential and a 
mass term of 
the form (with $X \equiv \Phi_0$, $Y \equiv \Phi_1$)
\begin{eqnarray}
\mathcal{W}(X,Y) &=& V(X) + t' Y^2 + cXY \nonumber \\ 
\mathcal{M}(X,Y) &=& 
   \left(\begin{array}{ll}
            \mathcal{M}_1(X) & 0 \\
             0 & 0 
          \end{array} \right)
\end{eqnarray}
by considering the following geometric deformation terms
\begin{eqnarray}
B(z, \omega) &=& \frac{1}{z} V(\omega) + \frac{t'}{z^3} \omega^2 + 
\frac{c}{2 z^2} \omega^2 \nonumber \\
M(z, \omega) &=& \frac{1}{z} \mathcal{M}_1(\omega)
\end{eqnarray}

\section{Solving the matrix model}

Matrix models of the type (\ref{Z0}) can be exactly solved. They are equivalent
to one--matrix models with a polynomial + logarithmic potential. In the 
following
we show that the technology of two--matrix models can be profitably used to 
solve them. This goes as follows. Let us consider, for simplicity, the case in
which in (\ref{Z0}) there is only one flavour (the extension to many flavours 
is straightforward). The integral over $\tilde q$ and $q$ can be explicitly
carried out and produces the determinant of ${\cal M}$ to the power $-1$. This
can be written as the exponent of $-\Tr \, \ln {\cal M}$ (for simplicity we set 
$g_s=1$, which is equivalent to rescaling the linear couplings in ${\cal W}$ 
and ${\cal M}$). Therefore (\ref{Z0}) is equivalent to
\begin{equation}
Z_0=\int dX 
{\rm e}^{\Tr\left( {\cal W}(X)- \ln {\cal M}(X) \right)}.\label{Z0log}
\end{equation}
where
\a
{\cal W}(X) = \sum_{k=1}^p t_k X^k, \qquad\qquad 
{\cal M}(X) = \sum_{k=1}^q s_k X^k.\label{potent}
\b

In order to be able to exploit the powerful technology of two--matrix models
we couple this model to a Gaussian one\footnote{Instead of a Gaussian model we 
could take any polynomial one, but the analysis would be much more complicated.} 
with a bilinear coupling (see section \ref{31} for a derivation of this model):
\begin{equation}
Z_0'=\int dX dY
{\rm e}^{\Tr \left( {\cal W}(X)- \ln {\cal M}(X) +c XY + t' Y^2 \right)}
\label{Z0'}
\end{equation}
After performing the path integral with the orthogonal polynomials method 
we will decouple the Gaussian model by setting $c=0$ (see, for instance, 
\cite{BBR}). By means of the orthogonal polynomials we can perform
the path integral explicitly and reduce the problem to that of solving the 
quantum equations of motion. Once this is done, the model is completely
solved because there exists a precise algorithm (based on the flow equations 
of the Toda lattice hierarchy) to calculate all the correlators.

Therefore the basic step in order to solve the model (\ref{Z0'}) is to 
solve the quantum equations of motion. In this case they take the form
\a
&&P^\circ(1)+ {\cal W}'(Q(1))+c Q(2)=
\frac {{\cal M}'(Q(1))}{{\cal M}(Q(1)) },\label{coupl1}\\
&& c Q(1)+2 t'Q(2)+\widetilde{\cal P}^\circ(2)=0,\label{coupl2}
\b
where a prime denotes functional derivative with respect to the (matrix) entry. 
$Q(1)$, $Q(2)$, $P^\circ(1)$, $\widetilde{\cal P}^\circ(2)$  
represent, in the basis of the orthogonal polynomials, multiplication by 
the eigenvalues $\lambda_1,\lambda_2$ of $X$ and $Y$, respectively, and the
derivatives with respect to the same parameters.
Eqs. (\ref{coupl1}, \ref{coupl2}) can be considered as the quantum analog of the 
classical equations of motion.
The difference with the classical equations of motion of the original matrix
model is that, instead of the $N \times N$ matrices $M_1$ and $M_2$, here we 
have infinite $Q(1)$ and $Q(2)$ matrices together with the quantum deformation 
terms represented by $P^\circ(1)$ and $\widetilde {\cal P}^\circ(2)$, respectively.
{}From the quantum equations of motion it follows that $Q(1)$ and $Q(2)$ are
Jacobi matrices, that is they have a
diagonal band structure and can be parameterized as follows.
\a
&&Q(1)=\sum_{i=0}^\infty(E_{i,i+1}+a_0(i)E_{i,i} + a_1(i+1)
E_{i+1,i})\label{Q1}\\
&&Q(2)=\sum_i \left(R(i+1)E_{i,i+1}+ \sum_{l=0}^{\infty} \frac{b_l(i)}{R(i+1)\ldots
R(i+l)}
E_{i,i+l}\right)\label{Q2}
\b
where $(E_{i,j})_{k,l}=\delta_{i,k}\delta_{j,l}$, 
$R({i+1}) \equiv h_{i+1}/h_i$.

To show something explicit without cluttering the paper with formulas, we
choose a simple example
\a
{\cal W}(X) = t_2 X^2 + t_3 X^3, \qquad\qquad {\cal M}(X) = s_0 + s_2 X^2
\label{poten}
\b
and write down explicit formulas only for the genus 0 case: this corresponds
to the lattice fields being independent of the lattice increment and to 
the replacement $n \to x=n/N$. Moreover we rescale the couplings as follows:
$t_2\to -N/(2g_s),t_3\to -Ng/g_s, c\to Nc, t'\to Nt'$ and define the ratio 
$s=s_2/s_0$. 
In the genus 0 case the above infinite
matrices are conveniently replaced by, \cite{BBR},
\a
&&Q(1)\to L=\zeta + a_0(x) + \frac {a_1(x)}{\zeta},\quad\quad
Q(2) \to \tilde L = \frac {R(x)}\zeta + \sum_{l=0}^\infty 
\frac {b_l(x)}{R^l(x)}\zeta^l \0\\
&&P^\circ(1)\to M = \frac x\zeta + {\cal O}(\zeta^{-2}),\qquad\qquad
\widetilde {\cal P}^\circ(2)\to \tilde M = \frac xR \zeta + {\cal O}(\zeta^{2})\0
\b
Eq.(\ref{coupl1},\ref{coupl2}) is intepreted as a set of infinitely many 
equations obtained
by expanding in $\zeta$ and equating the relevant coefficients.

The equations corresponding to (\ref{coupl2}) are easy to obtain. The crucial
ones are simply (see \cite{BBR})
\a
2 t'R +ca_1=0,\qquad 2 t' b_0 +c a_0 =0,\qquad x + 2 t' b_1 +
cR=0\label{coupl2'}
\b  
while the remaining ones determine the unknown elements of ${\cal P}^\circ(2)$.
{}From (\ref{coupl2'}) we can determine $R,b_0,b_1$ in terms of $a_0$ and $a_1$.
The result is as follows. Corresponding to $\zeta^q$, we have
\a
c\frac {b_q}{R^q} =3t_3  2 \sum_{k=0}^\infty \sum_{p=0}^{2k+1} (-1)^k s^{2k+1}
\frac{(2k+1)!}{(\frac {p-q}2)!(\frac {p+q}2)!(2k+1-p)!}a_0^{2k+1-p}a_1^{\frac
{p-q}2}, \label{R1}
\b
for $ q>1$. The sum over $p$ is limited to the $p$'s such that $p-q$ is even.
For $q<-1$ the equations determine the unknown components of 
$ P^\circ(1)$. For $q=1,0$ we have, respectively, after replacing
(\ref{coupl2'}),
\a
&&\frac {c^2a_1}{2t'} +a_1 +6 g a_0a_1 -x + 2s g_s \sum_{k=0}^\infty 
\sum_{l=0}^k (-1)^k s^{2k}
 \frac{(2k+1)!}{l!(l+1)!(2k-2l)!}a_0^{2(k-l)}a_1^{l+1}=0\label{R2}\\
&& \frac {c^2a_0}{2t'} +a_0 + 3 (a_0^2 g +2a_1)g + 2 s g_s
\sum_{k=0}^\infty  \sum_{2l=0}^{2k+1} (-1)^k s^{2k} 
\frac{(2k+1)!}{l!l!(2k+1-2l)!} a_0^{2k+1-2l}a_1^{l}=0.\label{R3}
\b
The equation corresponding to $\zeta^{-1}$ is a copy of (\ref{R2}) multiplied by
$a_1$. Now we can safely take the decoupling limit $c\to 0$. In this limit
$R\sim c$ and $b_l\sim c^{l+1}$, except for $b_1$ which remains finite.
The model splits into a Gaussian model we will henceforth forget about and the
cubic+logarithmic model we started with. The latter is determined by 
the following equations, obtained from (\ref{R2}, \ref{R3}),
\a
&&a_1 +6 g a_0a_1 -x + 2s g_s \sum_{k=0}^\infty 
\sum_{l=0}^k (-1)^k s^{2k}
 \frac{(2k+1)!}{l!(l+1)!(2k-2l)!}a_0^{2(k-l)}a_1^{l+1}=0\label{R2'}\\
&&a_0 + 3 (a_0^2 g +2a_1)g + 2 s g_s
\sum_{k=0}^\infty  \sum_{2l=0}^{2k+1} (-1)^k s^{2k} 
\frac{(2k+1)!}{l!l!(2k+1-2l)!} a_0^{2k+1-2l}a_1^{l}=0\label{R3'}
\b
These are the equations one has to solve in order to determine $a_0,a_1$ and
determine all the correlators. We notice that by setting $s=0$ one gets the
equations of the quantum Riemann surface of ref.\cite{BBR}, section 5.1.
Therefore, for $s$ small, (\ref{R2'},\ref{R3'}) can be thought to represent
a deformation of such Riemann surface.

It is not possible to find an exact compact solution of (\ref{R2'},\ref{R3'}), 
as in \cite{BBR}. However it is rather simple to find solutions in the form of
power series, which is just as well for the purpose of determining correlators.
For instance, for small $x$, we easily get
\a
&&a_0(x)= -\frac {6 g}{(1+2 s g_s)^2} x -36 \frac {9g^3-g_sgs^3-2g_s^2gs^4}
{(1+2 g_s s)^5} x^2+\ldots\label{sol1}\\
&&a_1(x)=  \frac 1{1+2 s g_s} x+6 \frac {6g^2 +g_s s^3 +2 g_s s^4}{(1+2 g_s
s)^5} x^2+\ldots\label{sol2}
\b

{}From such expansions we can compute the correlators of the operators
$\tau_k=\Tr(X^k)$ and $\sigma_k=\tilde q X^k q$. As for the former, everything
works exactly as in \cite{BBR}. They are obtained by extending the potential
as ${\cal W}(x)=\sum_{k=1}^\infty \hat t_k x^k$, differentiating the 
partition functions 
with respect to $\hat t_k$ and then setting $\hat t_k=0$ except for
$\hat t_2=t_2,\hat t_3=t_3$ (which we denote collectively by $\hat t =t$). 
The correlators
of the $\sigma_k$'s can instead be computed in the following way. Let us extend
the potential ${\cal M}(X)$ to $\hat{\cal M}(X) = \sum_{k=0}^\infty 
\hat s_k X^k$. After differentiating with respect to $\hat s_k$ we will set
the latter to zero except for $\hat s_0=s_0, \hat s_2=s_2$ (which we denote
collectively by $\hat s=s$). We obtain
\a
\left.\frac {\d Z}{\d\hat s_k}\right\vert_{\hat s=s} = 
\left.\int dX dq d\tilde q  \, 
\frac {\d e^{-S}}{\d\hat s_k}\right\vert_{\hat s=s}=
-\int dX \,\Tr \left(\frac {X^k}{{\cal M}(X)} \right) 
e^{\Tr \left( {\cal W}(X) -\ln {\cal
M}(x)\right)}  \label{sigmak}
\b
Expanding 
\a
\frac 1{{\cal M}(X)} = \frac 1{s_0} \left(1 -\frac ss_0 X^2 + (\frac
ss_0 X^2)^2-\ldots\right) \0
\b
in the integrand, we get
\a
<\sigma_n> =\left. -\frac 1s_0 \sum_{k=0}^\infty (-1)^k \left(\frac ss_0\right)^k  
\frac {\d Z}{\d t_{2k+n}}\right\vert_{\hat t=t}\label{tau2k+n}
\b
That is we can express the $\sigma_k$ correlators as series of the $\tau_k$ 
ones. 

\section{Conclusions}

In this letter we have analyzed topological open string theory
on a BPS sector of type IIB string theory on 
(partly resolved) non--compact Calabi--Yau manifolds and shown that D-branes 
placed on the blown--down cycles (singular points) produce flavour sectors. 
{}From the point of
view of gauge theory, the results we have obtained concern
the F-terms and therefore our analysis is completely blind to the dynamics 
of the gauge theory. To summarize,
our results amount in practice to the Dijkgraaf-Vafa \cite{DV} construction
with flavour.

Notice that our picture resembles very much the picture one obtains by
considering in type IIA 3 parallel NS 5-branes with D4 suspended between
them
\footnote{These 3 parallel NS 5-branes correspond to the $A_2$
case. More general cases should correspond to more complicated geometries.}.
Once a lateral 5-brane is moved to infinity, the related gauge sector
gets frozen and one is left with matter in the fundamental of the
remnant gauge symmetry with flavour symmetry \cite{Witten:1997sc}.
It would be quite interesting to figure out a detailed string duality 
connecting the IIA and IIB pictures. This problem was attacked long ago
in \cite{Hanany:1998it} where it was described as a T-duality.
More in general, it would be of some interest to extend our construction to
other partially resolved singular toric CY geometries and to be able to read
the blow--down operations on the cycles as actions on the related gauge theory
quiver diagrams.

The study of the geometric transitions, along the lines of \cite{DV,Diaconescu},
starting from partially resolved CYs
might also be of interest in the context of gauge-string correspondence and
could shed light on the gauge theory dynamics related to our construction. 
This goes however beyond the scope of this letter.

Finally let us stress that our construction in section 3 concerns
non--linear deformations of the complex structure of the space outside the
singularities, while for the singular part we limited ourselves to 
considering linear 
deformations (corresponding to the $\tilde Q Q$ terms in the effective 
superpotential).
It would be very interesting to know whether such deformations are 
implied by some 
geometrical constraint in varying the complex structures of singular spaces.
This issue might find a natural explanation in the much more elaborated 
framework considered in \cite{Aspinwall,AK}. 

\vspace{1 cm}
{\bf Acknowledgements}
We thank 
R.~Argurio,
M.~Bertolini,
U.~Bruzzo,
B.~Fantechi,
I.~Klebanov,
N.~Nekrasov,
A.~Tanzini and
A.~Tomasiello
for useful discussions. %

 This research has been supported by the Italian MIUR
under the program ``Teoria dei Campi, Superstringhe e Gravit\`a''.
The work of G.B. is supported by
the Marie Curie European Reintegration Grant MERG-CT-2004-516466,
the European Commission RTN Program MRTN-CT-2004-005104.

\vspace{1 cm}
\section*{Appendix. Partial $A_2$ resolution}

This Appendix is devoted to a nuts--and--bolts description
of the partial resolution of a surface with $A_2$ singularity.

Let us consider the $A_2$ singular surface given by the equation
\be
uv=y^3 \quad {\rm where}\quad (u,v,y)\in {\bf C}^3.\label{A2}
\ee
This surface has a doubly singular point at the origin in ${\bf C}^3$.

We construct the resolution of this singularity as two copies of 
${\cal O}(-2)_{{\bf P}^1}$ glued along two points by exchanging the base spaces 
with fibers. In formulas, this implies the overlapping conditions 
\be
z_i'=z_i^{-1}
\quad{\rm and}\quad
p_i'=z^2_i p_i\label{transf1}
\ee
on the intersections of two copies of north and south charts
$U_N^i\cap U_S^i$
with $i=1,2$;
the completely resolved space is then obtained by gluing the south patch 
$U_S^1$ and the north patch $U_N^2$ as
\be
z_2'=p_1
\quad{\rm and}\quad
p_2'=z_1.\label{transf2}
\ee

The blow-down map is given patch by patch by
\a
&&(u,v,y)=({p_1'}^2{z_1'}^3,p_1',p'_1z'_1) \quad{\rm on} \quad U_N^1\0\\
&&(u,v,y)=(p_1^2z_1,p_1z_1^2,p_1z_1) \quad{\rm on} \quad U_S^1
\label{complresol}\\
&&(u,v,y)=(z_2'^2p_2',p_2'^2z_2',p_2'z_2') \quad{\rm on} \quad U_N^2\0\\ 
&&(u,v,y)=(p_2,p_2^2z_2^2,p_2z_2) \quad{\rm on} \quad U_S^2\0
\b 
The counter-image of the origin in ${\bf C}^3$ consists of the 
union of the two ${\bf P}^1$s.  

The partial resolution of $A_2$ is then defined by forgetting one of the two 
copies of ${\cal O}(-2)_{\P1}$, say the one denoted by $i=2$, and replacing 
it with an additional fiber at the singular point.
The extra fiber can be recognized upon blow--down to be the fiber 
sitting at $z_2=0$, which is invisible to the other $\P1$.

Let us denote by $A_2'$ the partially resolved $A_2$ surface 
${\cal O}(-2)_{\P1} \vee {\bf C}$. $A_2'$ has a finite volume 2-cycle 
${\bf P}^1$ and a zero volume 2-cycle $pt$, 
the shrunk one, placed at the singular point. 

An explicit description of $A_2'$ could go as follows. Let us start from the 
completely resolved geometry (\ref{complresol}) above and define the new
coordinates
\be
\tilde u = p_1,\quad\quad \tilde v= p_1 z_1^2, \quad\quad \tilde y = p_1z_1
\label{A1coo}
\ee
in the chart $U^1_S$. In the other charts the new coordinates are given
by (\ref{transf1},\ref{transf2}). 

One gets
\be
\tilde u \tilde v= {\tilde y}^2\label{A1}
\ee
which is the equation of an $A_1$ singularity. The counterimage of the singular
point $\tilde u =\tilde v= {\tilde y}=0$ is the first sphere, parameterized by
$z_1$ and $p_1=0$. Therefore this cycle is squeezed to a point in the surface 
(\ref{A1}). The equations $\tilde v=\tilde y=0$ with 
$\tilde u$ generic represents the surviving cycle together with 
the fiber at $z_1'=0$. Eq.(\ref{A1}) represents therefore $A_2'$.
For completeness we can proceed to blowing down the second cycle via
\be
\hat u = {\tilde u}^2 \tilde v,\quad\quad \hat v =\tilde v,\quad\quad
\hat y = \tilde u\tilde v\label{A1'coo}
\ee
so that
\be
\hat u \hat v = {\hat y}^2\label{A1'}
\ee
The inverse image of the singular point $\hat u =\hat v= {\hat y}=0$ is 
precisely the second cycle. We remark that (\ref{A1'}), expressed in terms 
of the coordinates $u,v,y$, takes the form $y(y^3-uv)=0$, which represents
the reduced join of the surface (\ref{A2}) with the two original cycles plus
extra fibers.  

The same construction can be obtained in a neater, although somewhat pedantic 
way, via toric geometry.

\end{document}